\documentclass[10pt,conference]{IEEEtran}
\usepackage[utf8]{inputenc}
\usepackage{dsfont}
\usepackage{xcolor}
\usepackage{float}
\usepackage{enumitem}
\usepackage{amsmath}
\usepackage[normalem]{ulem}
\usepackage{siunitx}
\usepackage{graphicx}
\graphicspath{ {./images/} }
\usepackage{adjustbox}
\usepackage{dcolumn}
\usepackage{bm}
\usepackage[unicode=true]{hyperref}
\hypersetup{colorlinks=true,linkcolor=blue,filecolor=blue,urlcolor=blue,citecolor=blue}
\usepackage{nicefrac}
\usepackage{physics}
\usepackage[capitalize]{cleveref}
\usepackage[caption=false]{subfig}
\usepackage{authblk}
\usepackage{listings}
\usepackage{booktabs}

\definecolor{codegreen}{rgb}{0,0.6,0}
\definecolor{codegray}{rgb}{0.5,0.5,0.5}
\definecolor{codered}{rgb}{0.8,0,0}
\definecolor{backcolour}{rgb}{0.95,0.95,0.92}
\definecolor{commentcolor}{rgb}{0.2, 0.5, 0.6}

\usepackage{hyperref} 
\hypersetup{
    colorlinks=true,
    linkcolor=blue, 
    filecolor=blue,
    urlcolor=blue,
    citecolor=blue
}
\lstdefinestyle{mystyle}{
    backgroundcolor=\color{lightgrey},   
    commentstyle=\color{commentcolor}\itshape,
    keywordstyle=\color{codegreen}\bfseries,
    numberstyle=\tiny\color{codegray},
    stringstyle=\color{codered},
    basicstyle=\ttfamily\scriptsize,
    breakatwhitespace=true,         
    breaklines=true,                 
    captionpos=b,                    
    keepspaces=true,                 
    numbers=none,                    
    numbersep=5pt,                  
    showspaces=false,                
    showstringspaces=false,
    showtabs=false,                  
    tabsize=2,
    postbreak=\mbox{\textcolor{gray}{$\hookrightarrow$}\space},
    framexleftmargin=5pt,
    framextopmargin=0pt,
    framexbottommargin=0pt,
    frame=tb, framerule=0pt
}

\lstdefinelanguage{mypython}[]{Python}{
    morekeywords={as}
}

\lstdefinelanguage{XML}
{
  morestring=[b]",
  morecomment=[s]{<?}{?>},
  stringstyle=\color{codered},
  identifierstyle=\color{codegreen}\bfseries,
  keywordstyle=\color{blue},
  morekeywords={xmlns,version,type,name}
}

\definecolor{hyperlink_blue}{RGB}{0, 105, 171}
\definecolor{citation_blue}{RGB}{120, 2, 199}
\definecolor{url_blue}{RGB}{2, 28, 196}
\definecolor{lightgrey}{RGB}{240, 240, 240}

\usepackage{tikz}
\usetikzlibrary{arrows}

\newcommand{\CCC}[0]{$\mathds{C}^3$}

\begin{document}
\bstctlcite{IEEEexample:BSTcontrol}

\title{
Software tool-set for automated quantum system identification and device bring up 
}



\author[1, 2, 3]{Anurag Saha Roy}
\author[1, 2]{Kevin Pack}
\author[1, 2]{Nicolas Wittler}
\author[1, 3]{Shai Machnes}

\affil[1]{\small Peter Gr\"unberg Institute -- Quantum Computing Analytics (PGI 12), Forschungszentrum J\"ulich, D-52425 J\"ulich, Germany}
\affil[2]{\small Theoretical Physics, Saarland University, 66123 Saarbr\"ucken, Germany}
\affil[3]{\small Qruise GmbH, 66113, Saarbr\"ucken, Germany}

\maketitle


\begin{abstract}
	We present a software tool-set which combines the theoretical, optimal control view of quantum devices with the practical operation and characterization tasks required for quantum computing. In the same framework, we perform model-based simulations to create control schemes, calibrate these controls in a closed-loop with the device (or in this demo \textemdash by emulating the experimental process) and finally improve the system model through minimization of the mismatch between simulation and experiment, resulting in a digital twin of the device. The model based simulator is implemented using TensorFlow, for numeric efficiency, scalability and to make use of automatic differentiation, which enables gradient-based optimization for arbitrary models and control schemes. Optimizations are carried out with a collection of state-of-the-art algorithms originated in the field of machine learning. All of this comes with a user-friendly Qiskit interface, which allows end-users to easily simulate their quantum circuits on a high-fidelity differentiable physics simulator.
\end{abstract}


\section{Introduction}

Development of quantum technology devices, in particular quantum computers is an enormously difficult task. Focusing for the moment on quantum computation, one can estimate NISQ (Near-term Intermediate-Scale Quantum Computers, \cite{Preskill_2018_nisq}) devices become generally useful as we approach $100$ qubits. But to be able to use this number of qubits in a meaningful way, a quantum circuit must also have sufficient depth. For most useful applications \cite{bravyi2018quantum}\cite{beauregard2002circuit}\cite{suzuki2022computational}, a back of the envelope calculation thus suggests that an entangling gate error of $10^{-4}$ is an important step towards useful quantum computers, and progress towards that goal has been slow.

Whether aiming at error-corrected quantum computation or applications on NISQ devices, we posit one must significantly improve the methods by which we characterize quantum devices, and how we calibrate quantum controls.
One common approach, known as ad-hoc \cite{AdHOC}, consists of deriving controls in simulation and subsequent fine-tuning in experiment. This scenario leaves a lot to be desired: The need for calibration proves the model we have of the system is inaccurate, and the final calibrated pulses may perform well, but we do not know why, as we have no model which explains the changes in controls which have occurred during calibration. To make matters worse, we have not learned anything about the system in the process, and we are not in a better position to improve system design in the next hardware iteration.



To resolve this unsatisfactory situation, the authors have developed a novel methodology, \CCC\ -- Control, Calibration and Characterization \cite{wittler2021integrated}, which tightly integrates the calibration and characterization processes, to achieve better modeling of the system and hence better optimal control. This manuscript details the open-source (Apache 2.0) software implementation of this new methodology. And while there are several open-source software packages which perform either model-based optimal control, control calibration or limited system characterization \cite{Alexander_2020_QST,Ball_2020, QUA-libs, li2021qutip-qip, muller2021one}, to our knowledge \CCC\ is the first which combines all three in an integrated fashion, providing better insights into the sources of limitations in fidelities, and hence hints as to what should be addressed in the next hardware iteration.

The manuscript is organized as follows: We present the optimal control task of realizing a quantum computing gate-set on a single qubit in Section \ref{sec:oc}. To demonstrate the full \CCC\ procedure, we will consider two similar models, the \emph{simulation model}, to which we have full descriptive access, and the \emph{blackbox model}, which stands-in for a real experiment and thus only provides limited access by measurement. We also discuss the implementation of the underlying quantum simulation. In Section \ref{sec:calib} we simulate the procedure of calibrating the gate-set by taking measurements on the \emph{blackbox}. On the basis of the data taken during calibration, we perform model learning in Section \ref{sec:learning}, by simulating the calibration experiments on the \emph{simulation model} to bring the two to agreement. Finally, we discuss the Qiskit \cite{Qiskit} interface in Section \ref{sec:qiskit} and give an overview of the software architecture in Section \ref{sec:arch}.

As a companion piece, an example Jupyter notebook that contains the code and reproduces the plots in this manuscript is accessible through this link - \href{https://bit.ly/c3-full-example}{bit.ly/c3-full-example}. This is released as part of the source code on Github at   \href{https://github.com/q-optimize/c3}{q-optimize/c3}.

\section{Optimal Control}\label{sec:oc}

When employing optimal control techniques in the practical operation of quantum devices, it is desirable to have access to a model of the device for analysis and the derivation of advanced control schemes. From a numerical perspective, algorithms that make use of the gradient of the goal function, e.g. L-BFGS \cite{L-BFGS}, can speed up  optimizations significantly compared to gradient-free algorithms \cite{Machnes18}. Access to gradients of a quantum dynamics simulation is not trivial and thus procedures such as Krotov \cite{Koch-Krotov-Main}, GRAPE \cite{GRAPE} and GOAT \cite{Machnes18} require a specific formulation of the control problem in goal function and system descriptions. In the approach presented here, we make use of the numerics package Tensorflow, developed for machine learning applications, to provide both high compute performance and the access to automatic differentiation which lifts most limitations in formulating the problem.

\subsection{The problem statement}
For demonstration purposes, we will concern ourselves in this paper with the realization of a gate-set of $\pi/2$ rotations around the $x$ and $y$ axis of the Bloch sphere of a single qubit. This system is described by the Hamiltonian
\begin{equation}\label{eq:ham}
		H(t)/\hbar =
		\omega b^\dagger b - \frac{\delta}{2}
		\qty(b^\dagger b - 1) b^\dagger b 
		+ c(t) \qty(b+b^\dagger) \ ,
\end{equation}
where $b$ and $b^\dagger$ are the bosonic ladder operators, $\omega$ and $\delta$ the qubit frequency and anharmonicity and $c(t)$ the time-dependent control field.
 
The following snippet creates the qubit as an object in the \CCC\ package.
 \begin{lstlisting}[language=Python]
 q1 = Qubit(
	 name="Q1",
	 desc="Qubit 1",
	 freq=Quantity(
		 value=freq,
		 min_val=4.995e9,
		 max_val=5.005e9,
		 unit="Hz 2pi",
	 ),
	 anhar=Quantity(
		 value=anhar,
		 min_val=-380e6,
		 max_val=-120e6,
		 unit="Hz 2pi",
	 ),
 )
\end{lstlisting}
Properties that will be optimized are instantiated from the \texttt{Quantity} class, which handles physical units and limits.
Similarly, the drive is created by
 \begin{lstlisting}[language=Python]
	drive = Drive(
	name="d1",
	desc="Drive 1",
	comment="Drive line 1 on qubit 1",
	connected=["Q1"],
	hamiltonian_func=hamiltonians.x_drive,
)
\end{lstlisting}
where the \texttt{connected} parameter indicates this drive acts on the qubit created above. When the model is created with \texttt{model = Model([q1], [drive])}, the matrix representation of the operators in (\ref{eq:ham}) are created and stored for later use. The operation $U$ on the qubit, affected by the control field is obtained by the formal solution of the Schrödinger equation
 \begin{equation}
 	U(T) = \exp(-\frac{i}{\hbar}\int_{0}^{T}H(t) d t) \, .
 \end{equation}

 The control task now consists of finding the control fields $c(t)$ which implement the gate-set
\begin{equation}
		\mathcal{G} = \qty{X_{\pi/2}, Y_{\pi/2}, X_{-\pi/2}, Y_{-\pi/2} } 
		\label{eq:gates} \ ,
\end{equation}
of rotations of the qubit spin around the axis of the Bloch sphere. To determine the fidelity of a single operation, we use the overlap
\begin{equation}
    \mathcal{F}[U(t)] = \abs{\frac{1}{\dim U}\Tr{U^\dagger(t)U_\text{ideal}}}^2
\end{equation}
which we average over all operations of the gate-set to compute the goal function value.

\subsection{Signal generation}
\begin{figure}[t]
	\begin{tikzpicture}
		\path (0,2) node[draw,double,rounded corners](x1) {$\varepsilon(t)$}
		-- (0,0) node[draw,double,rounded corners](x2) {$u(t)$} node[left, midway]{AWG}
		node[right, midway]{sampling}
		-- (0,-2) node[draw,double,rounded corners](x3) {$c(t)$} node[right, midway]{function} node[left, midway]{transfer}
		-- (5 ,-2) node[draw,double,rounded corners](x4) {$\ket{\psi(t)}=U(t)\ket{\psi(0)}$} node[above, midway]{time evolution \phantom{fooooo}}
		-- (5,2) node[draw,double,rounded corners](x5) {$\{p_i\}$} node[left, midway]{readout};
		\draw[->, thick] (x1) -- (x2);
		\draw[->, thick] (x2) -- (x3);
		\draw[->, thick] (x3) -- (x4);
		\draw[->, thick] (x4) -- (x5);
	\end{tikzpicture}
	\caption{Figure and caption adapted from \cite{wittler2021integrated}: The process of simulating experimental procedure for signal processing and readout. 
		The control function is specified by some function $\varepsilon$ and specifies the line voltage $u(t)$
		by an arbitrary waveform generator (AWG) with limited bandwidth.
		Electrical properties of the setup, such as impedances, are expressed as a line transfer
		function $\varphi$, resulting in a control field $c(t)=\varphi\qty[u(t)]$, as in Eq.
		(\ref{eq:ham}).
	}
	\label{fig:flow-exp-to-sim}
\end{figure}
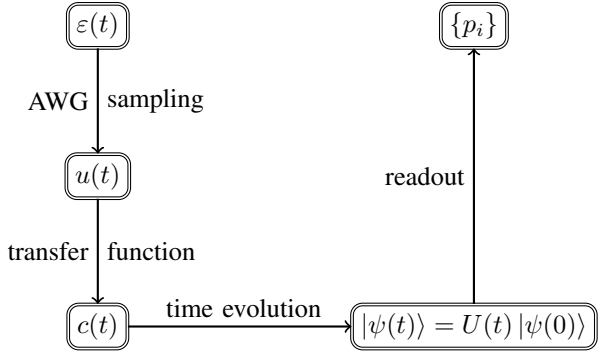

The control field $c(t)$, as seen by the qubit and written in the Hamiltonian, is typically different from the programmed field in signal generators of an experimental setting. The mathematical background is described in \cite{wittler2021integrated}. The process is implemented as a generator using a number of devices to realize the signal chain.
 \begin{lstlisting}[language=Python]
generator = Generator(
	devices={
		"LO": ...,
		"AWG":...,
		"DigitalToAnalog" ...,
		"Mixer": ...,
		"VoltsToHertz": ...,
	},
	chains={
		"d1": {
			"LO": [],
			"AWG": [],
			"DigitalToAnalog": ["AWG"],
			"Mixer": ["LO", "DigitalToAnalog"],
			"VoltsToHertz": ["Mixer"],}},)
\end{lstlisting}
In the generator, a number of devices are arranged into signal chains by specifying their inputs and finally, their voltage signals are converted to energy (or frequency) scale to be put into the Hamiltonian.

\subsection{Simulating quantum dynamics}
The simulation of quantum dynamics is at the core of both the optimal control and model learning applications. To efficiently simulate a sequence of gates in the gate-set in a Markovian context, it makes sense to compute the Unitary representations of the gate-set only once and multiply them as needed to create the sequence.  We employ a technique similar to the virtual Z \cite{mckay2017efficient} phenomenon by multiplying the unitary representations of each gate by $\exp(\omega_d T b^\dagger b)$ which realigns the $x,y$ axis of the qubit and the drive field.

\subsection{Optimization}

To manipulate the control parameters, we use the \texttt{ParameterMap} that contains methods to collect optimizable parameters from both model and control components. In the example, we parametrize the control function with three parameters, the amplitude of the pulse, the DRAG parameter \texttt{delta}, the \texttt{frequency} offset which modulates the resonance and \texttt{framechange}, a phase which allows fine-tuning of the rotation. 

The optimization is then carried out by the module
\begin{lstlisting}[language=Python]
opt = OptimalControl(
	dir_path=log_dir,
	fid_func=unitary_infid_set,
	fid_subspace=["Q1"],
	pmap=parameter_map,
	algorithm=lbfgs,
	options={"maxfun" : 150},
	run_name="better_X90"
)
\end{lstlisting}
by specifying a fidelity function, including the subspace in the case of larger Hilbert space. Both the fidelity and the algorithm can be chosen from a library included in the package or from a custom user implementation with the same signature.

\section{Calibration}\label{sec:calib}
Having created an initial starting pulse through the usage of the tools in the optimization phase, we now continue with the refinement of said pulse in a closed feedback loop, the calibration step.

\subsection{Automated Calibration}
In traditional calibration, a portfolio of experiments executed in a specific sequence is used to gain knowledge of the different parameters of the quantum mechanical system and control pulse necessary to implement a desired control. This obviously does not scale to 100s of qubits. Henceforth, we advocate an automated approach to calibration where we try to minimize a goal function, which acts as a figure of merit for the desired operation. This approach is by no means new, and has been used successfully in the past to fine tune control pulses as can be seen in \cite{Werninghaus_2021}.

In this example, we decided to use ORBIT \cite{Kelly2014} as a goal function to minimize, and in the following chapter it will be shown how an ORBIT experiment can be set up in the \CCC~code. The task of calibration itself can be further simplified as a black box optimization, since we have to assume that our initial model of the system is not accurate. Here, we use CMA-ES, an advanced gradient free optimization algorithm. 

\subsection{ORBIT as Loss Function}

The creation of the ORBIT sequence in \CCC~is done through a helper function, namely \texttt{single\_length\_RB}, contained in the \texttt{qt\_utils} module. 
\begin{lstlisting}[language=Python]
qt_utils.single_length_RB(
        RB_number=1, RB_length=5, target=0
    )
\end{lstlisting}
The function has access to an internal list of decompositions of Clifford gates made up of available elemental operations. In this case, these elemental operations are the gates specified in $\mathcal{G}$ given in Eq. \ref{eq:gates}. Which elemental operations are available usually depends on the experimental hardware. The function then returns a list of instruction names corresponding to the previously defined available instructions, which can be sent to the simulation as in this example or to a real quantum computer. We use the mean over the population probabilities of the desired final state as the final result. This concludes the evaluation of the ORBIT sequences for the simulation. 

\subsection{Setup of The Optimization Algorithm}
After the preparation of the loss function, we continue with the setup of the optimizer algorithm. We rely for the minimization of the loss function on the covariance matrix adaptation evolution strategy or CMA-ES algorithm provided by the python package \texttt{pycma} \cite{pycma}. Several relevant hyperparameters for the optimizer are also set up at this stage, the detailed description of which is beyond the scope of this manuscript. 

\begin{lstlisting}[language=Python]
alg_options = {
    "popsize" : 10,
    "maxfevals" : 300,
    "init_point" : "True",
    "tolfun" : 0.01,
    "spread" : 0.1
  }
\end{lstlisting}

\subsection{The Calibration Class}
After all preparations are done, we can finally continue by setting up the actual calibration object, which takes the previously created ORBIT loss function and the hyperparameters of the optimizer as input. 

\begin{lstlisting}[language=Python]
# Create a temporary directory to store logfiles, modify as needed
log_dir = "c3example_calibration"

opt = Calibration(
    dir_path=log_dir,
    run_name="ORBIT_cal",
    eval_func=ORBIT_wrapper,
    pmap=parameter_map,
    exp_right=simulation,
    algorithm=cmaes,
    options=alg_options
)
\end{lstlisting}

\begin{figure}[ht!]
        \centering
        \includegraphics[width=\linewidth, trim=0 0 3em 0]{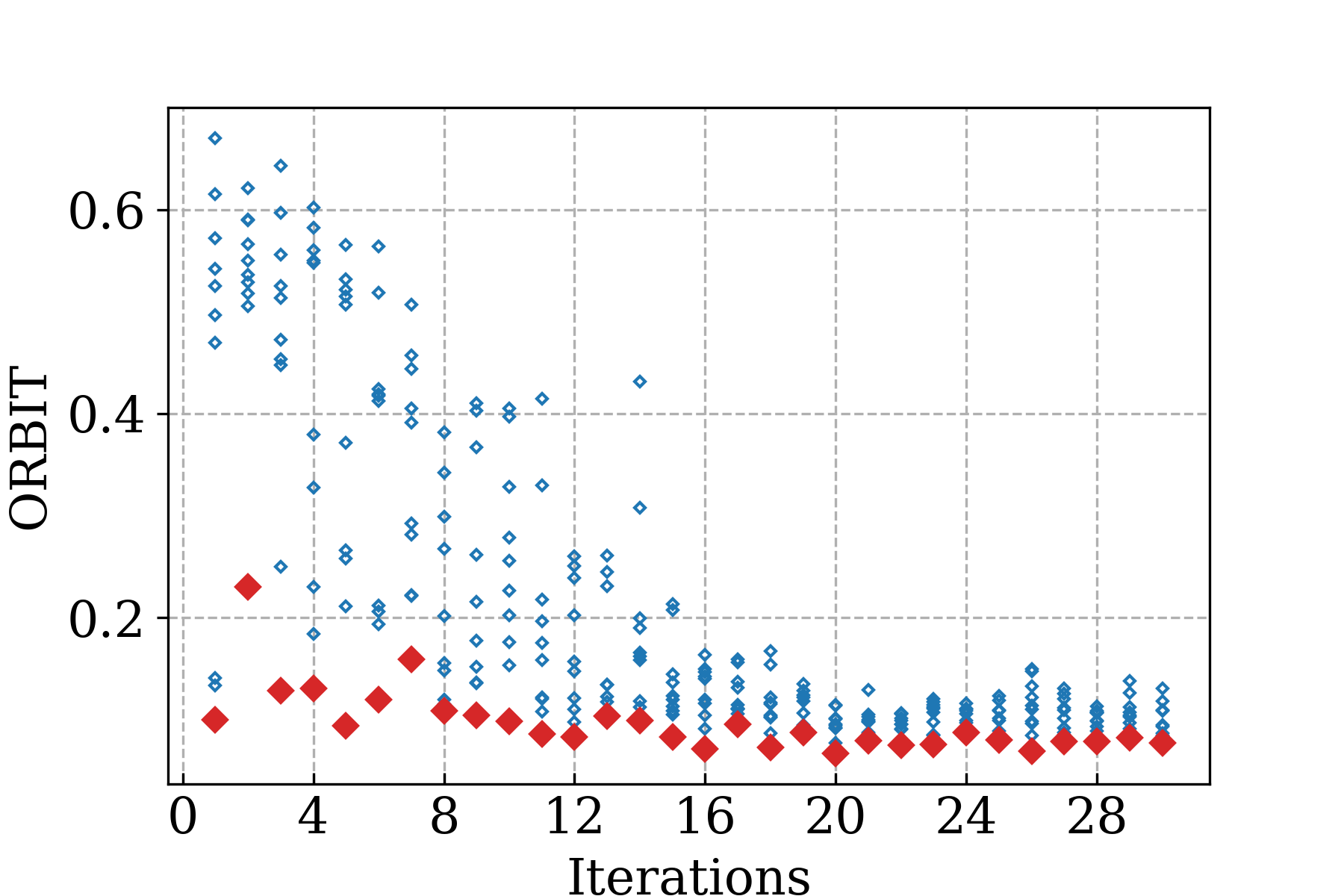}
        \caption[Calibration]{The results of the ORBIT evaluation as function of the iterations of the CMA-ES algorithm. The light blue diamond markers represent the function value of each point in a generation, while the red diamond marker is used to emphasize the best set of parameters in each generation.}
        \label{fig:orbit_calibration}
\end{figure}

With the final preparations completed, we can now begin the calibration process simply by calling the method \texttt{optimize\_controls} of the calibration class:
\begin{lstlisting}[language=Python]
opt.optimize_controls()
\end{lstlisting}
Fig. \ref{fig:orbit_calibration} shows the final result of the calibration step. We can see how the values of the loss function decrease indicating a successful calibration, which results in an improvement in the final gate fidelity.
It is noteworthy that the return value of the ORBIT loss function is not a gate fidelity or infidelity. This means, that any improvement of the goal function cannot be directly expressed as improvement in gate fidelity. Hence, even small improvements in the goal function can lead to big improvements in the gate fidelity.

\section{Model Learning}\label{sec:learning}

Model based Optimal Control can only be as good as the model, which means there is always scope for further refining the parameters of your model and also enriching your existing model with additional components. An intuitive approach to solving the former challenge (of tuning your model parameters to better represent reality) is to use Machine Learning techniques for extracting the model parameters from experimental data. 
\subsection{The Dataset}
The dataset for learning model parameters can be generated from any experiment that can be reproduced in simulation. For this example, we consider the data (stored a key-value pairs) from the ORBIT Calibration step, during which we obtained pairs of gate sequences and corresponding measurement results. 

\begin{table}[ht!]
\centering
\caption[ModelLearning]{Dataset for Model Learning}
\label{tab:dataset}
\begin{tabular}{lllll}
\hline
params & seqs & results & results\_std & shots \\ \hline
\begin{tabular}[c]{@{}l@{}}{[}376.655 mV, \\ -954.644 m, \\ -50.333 MHz 2pi, \\ -1.479 mrad{]}\end{tabular} &
  \begin{tabular}[c]{@{}l@{}}{[}{[}rx90p\ldots{]},\\ {[}ry90p\ldots{]},\\ \vdots \\ {[}rx90m\ldots{]}{]}\end{tabular} &
  \begin{tabular}[c]{@{}l@{}}{[}0.1421,\\  0.1606,\\ \vdots\\ 0.2507{]}\end{tabular} &
  \begin{tabular}[c]{@{}l@{}}{[}0.0137,\\ 0.0129,\\ \vdots \\ 0.0108{]}\end{tabular} &
  \begin{tabular}[c]{@{}l@{}}{[}1000, \\ 1000, \\ \vdots \\ 1000{]}\end{tabular} \\
       &      &         &              &       \\
       &      &         &              &       \\ \hline
\end{tabular}
\end{table}

In Table \ref{tab:dataset}, \texttt{params} is the parameters of the pulse during that specific run and \texttt{seqs} stores the exact ORBIT sequence that was used; while \texttt{results}, \texttt{results\_std} and \texttt{shots} tell us about the measurement outcome, standard deviation and number of shots respectively.
The dataset also has the required metadata, such as the \texttt{opt\_map} which contains information about the specific pulse parameters being optimized during that calibration run.

\subsection{Model and Loss Function}

We use the same model as described in the Optimal Control section since the goal of this step is to fine tune the model parameters by learning from the calibration data. We simulate the ORBIT sequence as contained in the dataset using this model and then compare the measurement outcomes of the simulation and the experiment. This comparison takes place in the form of a loss function\cite{wittler2021integrated} which is outlined in Eq. \ref{eq:loglikelihood}:
\begin{equation}\label{eq:loglikelihood}
    f_3(\beta) = f_\text{LL}\qty(\mathcal{D} \vert \beta) =
        \frac{1}{2N}\sum_{n=1}^{N}\qty[\qty(\frac{m_n-\widetilde{m}_n}{\widetilde{\sigma}_n})^2-1]
\end{equation}
that captures how well the model prediction $\widetilde{m}_{n}$, with standard deviation $\widetilde{\sigma}_n$, agrees with the recorded values $m_{n}$, in the dataset $\mathcal{D}$ for the model parameter values $\beta=(\omega_i, \delta_i, ...)$.

For the purpose of learning, we choose a subset of the entire data since simulating the whole dataset would be a significant computational challenge, as shown in the code snippet with accompanying explanatory comments.

\begin{lstlisting}[language=python]
datafiles = {"orbit": DATAFILE_PATH} # path to the dataset
run_name = "simple_model_learning" # name of the optimization run
dir_path = "ml_logs" # path to save the learning logs
algorithm = "cma_pre_lbfgs" # algorithm for learning
options = {
    "cmaes": {
        "popsize": 12,
        "init_point": "True",
        "stop_at_convergence": 10,
        "ftarget": 4,
        "spread": 0.05,
        "stop_at_sigma": 0.01,
    },
    "lbfgs": {"maxfun": 50, "disp": 0},
} # options for the algorithms
sampling = "high_std" # how data points are chosen from the total dataset
batch_sizes = {"orbit": 2} # how many data points are chosen for learning
state_labels = {
    "orbit": [[1,],[2,],]
} # the excited states of the qubit model, in this case it is 3-level
\end{lstlisting}

To start the Model Learning process, an object of the relevant class is instantiated with the parameters defined in the previous code snippet, followed by invoking the canonical \texttt{run()} function on this object.

\begin{lstlisting}[language=python]
opt = ModelLearning(
    datafiles=datafiles,
    run_name=run_name,
    dir_path=dir_path,
    algorithm=algorithm,
    options=options,
    sampling=sampling,
    batch_sizes=batch_sizes,
    state_labels=state_labels,
    pmap=parameter_map,
)
opt.set_exp(simulation)
opt.run()
\end{lstlisting}
An interesting aspect to note here is the choice of the \texttt{cma\_pre\_lbfgs} algorithm for optimization. This algorithm is a mix of gradient-free (blackbox) global optimization and gradient-based local optimization. The CMA-ES undertakes a global optimization run and once it is approximately close to the global minima region, it switches over to a gradient based L-BFGS run to identify the final minima. 
\begin{figure}[ht!]
        \centering
        \includegraphics[width=0.95\linewidth]{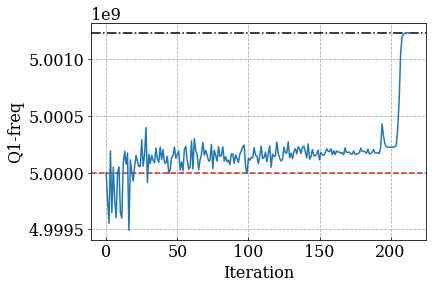}
        \caption{The Model Learning process for the Qubit Frequency (\texttt{Q1-freq}) \textemdash showing the initial search with CMA-ES followed by L-BFGS \textemdash as it converges to the true value of $5.001$ GHz after starting from the assumed value of $5$ GHz in approximately 200 iterations of the learning step. }
        \label{fig:model_learning}
\end{figure}

\section{Qiskit Interface}\label{sec:qiskit}
An important aspect of this software library is its intuitive and flexible interface which  is achieved with the help of the Qiskit interface, the implementation of which is discussed below.

\subsection{The \texttt{qiskit} backend in \texttt{c3-toolset}}
This library includes a \texttt{c3.qiskit} submodule which allows users to have a drop-in replacement for running their existing qiskit circuits on the high-fidelity \texttt{c3-toolset} physics simulator. A minimum example for this is below:

\begin{lstlisting}[language=python]
qc = QuantumCircuit(1)
qc.append(RX90pGate(), [0])
c3_provider = C3Provider()
c3_backend = c3_provider.get_backend("c3_qasm_physics_simulator")
c3_backend.set_c3_experiment(simulation)
c3_job_unopt = c3_backend.run(qc)
result_unopt = c3_job_unopt.result()
res_pops_unopt = result_unopt.data()["state_pops"]
\end{lstlisting}

\begin{figure}[ht!]
        \centering
        \includegraphics[width=0.95\linewidth]{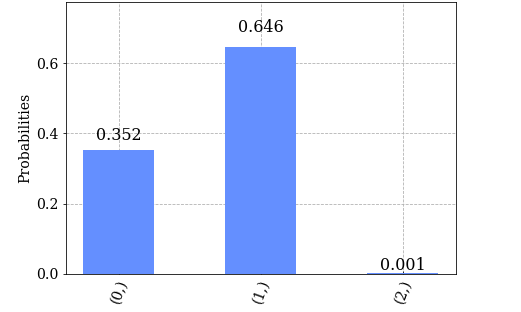}
        \caption{Results of running a qiskit circuit on the \texttt{c3-toolset} physics simulator with an unoptimized gateset.}
        \label{fig:qiskit}
\end{figure}

The histograms obtained from running this circuit, as shown in Fig. \ref{fig:qiskit} reflect the un-optimized nature of the gates, along with some leakage to higher levels of the transmon. The same circuit when run after the Optimal Control step, returns result populations much closer to ideally expected values.

\subsection{Pulse Level Control}
The qiskit interface also gives more advanced users complete pulse level control, allowing them to execute routines such as pulse-level noise mitigation or replicate complete calibration runs on the digital twin. This pulse level access is provided through two different interfaces. Firstly, the user can supply additional options to the qiskit \texttt{backend.run()} call, which allows them to define various pulse parameters, such as the amplitude, frequency, detuning etc specific to this particular circuit. Otherwise, users can also provide a completely arbitrary piece-wise constant waveform as pairs of In-phase and Quadrature values; as a parameter for custom user-defined gates.

\section{Software Architecture}\label{sec:arch}

The \texttt{c3-toolset} software is based on three main building blocks (\texttt{Model}, \texttt{Generator}, \texttt{Instructions}) that form the foundation of all the modelling and calibration tasks, and depending on the use-case, some or all of these blocks might be useful, together with the tools necessary to perform optimization tasks. 
\subsection{Model}

A theoretical Physics-based model of the Quantum Processing Unit. This
is encapsulated by the \texttt{Model} class which consists of objects
from the \texttt{chip} and \texttt{tasks} library. \texttt{chip}
contains Hamiltonian models of different kinds of qubit realisations,
along with their couplings while \texttt{tasks} let you perform common
operations such as qubit initialisation or readout. 

\subsection{Generator}

A digital twin of the electronic control stack associated with the
Quantum Processing Unit. The \texttt{Generator} class contains the
required encapsulation in the form of \texttt{devices} which help model
the behaviour of the classical control electronics taking account of
their imperfections and physical realisations.

\subsection{Instruction}

Once there is a software model for the QPU and the control electronics,
one would need to define Instructions or operations to be perform on
this device. Pulse shapes are described through a \texttt{Envelope} along 
with a \texttt{Carrier}, which are then wrapped up in the form of 
\texttt{Instruction} objects.

\subsection{Parameter Map}
The \texttt{ParameterMap} helps to obtain an optimizable vector of
parameters from the various theoretical models previously defined. This
allows for a simple interface to the optimization algorithms which are
tasked with optimizing different sets of variables used to define some
entity, e.g, optimizing pulse parameters by calibrating on hardware or
providing an optimal gate-set through model-based quantum control.

\subsection{Experiment}

With the building blocks in place, we can bring them all together
through an \texttt{Experiment} object that encapsulates the device
model, the control signals, the instructions and the parameter map. Note
that depending on the use only some of the blocks are essential when
building the experiment.

\subsection{Optimizers}

At its core, \texttt{c3-toolset} is an optimization framework and all of
the three steps - Open-Loop, Calibration and Model Learning can be
defined as a optimization task. The \texttt{optimizers} contain classes
that provide helpful encapsulation for these steps. 

\subsection{Libraries}

The \texttt{c3.libraries} sub-module includes various helpful library of
components that are used somewhat like Lego pieces when building the
bigger blocks, e.g, \texttt{hamiltonians} for the \texttt{chip} present
in the \texttt{Model} or \texttt{envelopes} defining a control
\texttt{pulse}.

\section{Outlook}

We have described \CCC, an open-source software package for the characterization, control planning and calibration for quantum computers and other quantum technology devices. 

Since publishing the initial paper \cite{wittler2021integrated}, significant developments have been undertaken by a number of contributors on \href{https://github.com/q-optimize/c3}{github.com/q-optimize/c3} to facilitate application in several projects and physical platforms. Development is ongoing, focusing on improving performance and flexibility. In the near future we are planning on adding support for NV-centers, Rydberg atoms and trapped ion systems. \CCC\, will also be the basis on which additional capabilities will be constructed. Co-design will allow one to simultaneously optimize model and control parameters, to find the model which will, when fabricated, provide the best performance. Sensitivity analysis can report on the distance between the model and the data in terms of standard deviations, and how this distance changes as a function of varying values of model parameters. Experiment Design will allow us to plan which data must be taken, i.e. which experiments must be performed. in order to efficiently reduce error bars of a specified model parameter. Finally, we note that much of what has been described in the \CCC\, methodology is not actually specific to quantum systems, and may be adapted to other realms of physics using the suitable digital twin. 


\section*{Acknowledgements}
This work was supported by the European Commission through the OpenSuperQ project (Grant Nr. 820363), by the Germany Ministry of Science and Education (BMBF) through projects VERTICONS and DAQC (Grant Nr. 13N14872 and 13N15688) and the Helmholtz Validation Fund project "Qruise" (HVF-00096).

\bibliographystyle{IEEEtran}
\bibliography{bibliography.bib}

\end{document}